\documentclass[aps,twocolumn,eqsecnum,showpacs]{revtex4}

\usepackage{dcolumn}
\usepackage{amsmath}
\usepackage{graphics}

\begin{document}
\title{Near-resonance light scattering from a high-density, ultracold atomic $^{87}$Rb Gas}

\author{S. Balik, A.L. Win and M.D. Havey}
\affiliation{Department of Physics, Old Dominion University,
Norfolk, VA 23529}
\date{\today }
\email{mhavey@odu.edu}

\author{I.M. Sokolov and D.V. Kupriyanov}
\affiliation{Department of Theoretical Physics, State Polytechnic
University, 195251, St.-Petersburg, Russia}%

\date{\today }

\begin{abstract}
We report a combined experimental and theoretical investigation of
near resonance light scattering from a high-density and
ultracold atomic $^{87}$Rb gas.  The atomic sample, having a peak
density $\sim 5\cdot10^{13}$ atoms/cm$^{3}$, temperature $\sim$ 65
$\mu$K and initially prepared in the F = 1 lower energy $^{87}$Rb
hyperfine component, is optically pumped to the higher energy F =
2 hyperfine level. Measurements are made of the transient
hyperfine pumping process and of the time evolution of scattering
of near resonance probe radiation on the $F = 2 \rightarrow F' =
3$ transition. Features of the density, detuning, and temporal dependence of the signals are attributed to the high
density and consequent large optical depth of the samples.
\end{abstract}


\maketitle%
\section{Introduction}
Development of techniques to cool and trap atomic gases
\cite{Metcalf,Grimm} has revolutionized many traditional atomic
physics research areas and, at the same time, has stimulated new
connections and types of interdisciplinary specialization. In atomic physics, original efforts were primarily directed towards observation and research on Bose-Einstein Condensation \cite{Pethick,Giorgini,Pitaevskii}. In turn, this has stimulated a vast amount of research in a wide range of areas including quantum information science
\cite{Bouwmeester,Lukin}, quantum optics
\cite{Milonni,Marangos,Hau,Braje}, precision measurements
\cite{Jin,Campbell,Ye}, plasma physics \cite{Rolston,Killian}, and
molecular spectroscopy \cite{Carr,Weidemuller}. These areas, and others, have been
transformed by the combination of ultracold experimental facility
and theoretical understanding of the physical processes. Among the
important characteristics of utilized atomic gases are the gas
temperature, density, and spin polarization.  For example, storage
of individual photon wave packets in ultracold gases combines
quantum optical techniques of coherent dark state formation and
electromagnetically induced transparency to address a critical
area of quantum information processing \cite{Lukin1,Kuzmich}. The
lifetime of the atomic spin wave, which determines the storage
time of the photonic information in the form of a dark state
polariton, depends on the gas temperature, collision rates, and
local magnetic environment.  In another research area,
photoassociative formation of ultracold diatomic molecules depends
quadratically on the density of the parent ultracold atomic gas.
Likewise, by initiating formation of an ultracold plasma in a high
density atomic gas, one can attain strong coupling with sufficient
ionization of the gas.

Over the past decade, a number of research groups
\cite{Havey,LabeyrieReview,Akkermans0,LPLReview,CommentAMO,KaiserHavey,Muller}
have concentrated on near-resonance light scattering in ultracold
atomic gases. A particular interest has been coherent multiple
scattering of light when the influence of atomic motion is of
minimal importance. Then the natural length scale for the photon
multiple scattering is the optical mean-free-path, given by
$\emph{l}$ = $1/\rho \sigma$, where $\rho$ is the atomic density
and $\sigma$ the cross-section for atomic light scattering in a
weak field. For light of wave vector magnitude $k = 2\pi/\lambda$,
a useful dimensionless parameter is the product k$\emph{l}$ \cite{Ioffe}. When
k$\emph{l}$ $>>$ 1, the so called weak localization regime
\cite{Akkermans0}, light scattering can be thought of as a
sequence of scattering and propagation events \cite{LPLReview}. In
this regime, an observable that survives configuration averaging
is the coherent backscattering cone, which is a few
milliradian-width feature that displays an enhancement of as much
as a factor of two over the incoherent albedo for back scattered
light \cite{Ishimaru,Wolf,LTBMMK}. In ultracold atomic gases, the
angular shape and peak enhancement of this spectral feature has
been studied for a wide range of conditions, including spectral
detuning from resonance, light polarization, probe light intensity
\cite{Mollow}, sample size and optical density, and external
magnetic field
\cite{Kaiser2,Strontium,Kaiser3,Kaiser4,Kaiser5,Antilocalization,KulatungaCBS,KupriyanovCBS,Strongfield}.
Observation of the coherent backscattering cone established that
weak-field multiple light scattering, even for resonance radiation, is a
coherent process. When the atomic density is significantly
increased, so that the parameter k$\emph{l}$ $\sim$ 1 (the
so-called Ioffe-Regel condition \cite{Ioffe}), there are a number
of atoms within a volume $1/k^{3}$, and light scattering
\cite{Tannoudji} becomes a cooperative process
\cite{KupriyanovLightStrong,Akkermans0,Cooperative,Akkermans1,Bienaime1,Bachelard,Bienaime2,Bender,Bux}. In this
regime, it is expected that a number of fascinating quantum
optical processes, including Anderson localization of light
\cite{Anderson,Wiersma1,Chabanov1,Maret1,Maret}, and atom-based random lasing
\cite{Cao,Wiersma,Conti,KaiserRandom1,KaiserRandom2}, may emerge.
It is these processes, and the conditions under which they might
be experimentally studied, that are the main motivations for the
current research program.

In the present paper, we describe our experimental approach to
obtaining atomic conditions, and particularly atomic densities, in
the regime where k$\emph{l}$ $\sim$ 1.  We also describe the
method we use to prepare the atomic sample so that large orders of
multiple scattering may be obtained, along with the associated
experimental observations and their interpretation.  This is
followed by presentation and discussion of experimental results of
the density and detuning dependence of time-dependent light
scattering on the nearly-closed F = 2 $\rightarrow$ F$^{\prime}$ =
3 hyperfine component of the $^{87}$Rb D2 transition.  The experimental results are also compared with those obtained by rigorous theoretical treatment \cite{KupriyanovLightStrong,JETP1} of the dynamics under conditions similar to those realized in the experiments.

\section{Experimental Approach}

\begin{figure}[htpb]
\includegraphics{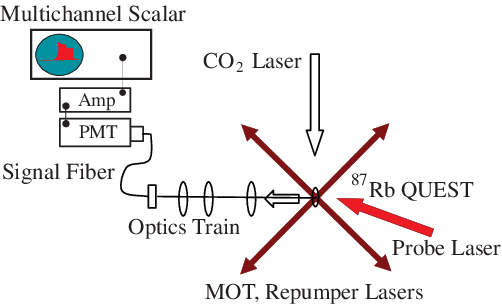}%
\caption{Schematic drawing of the experimental apparatus. In the
figure QUEST stands for quasi electrostatic trap and PMT refers to a
photomultiplier tube. MOT refers to a magneto-optical trap, while $CO_{2}$ laser
indicates a carbon-dioxide laser. Drawing not to scale.}
\label{fig1}%
\end{figure}

\begin{figure}[htpb]
\includegraphics{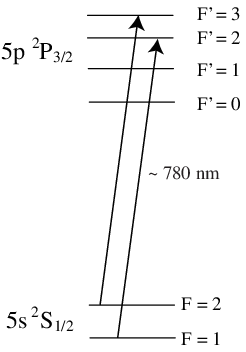}%
\caption{Schematic energy level diagram for $^{87}$Rb showing the
optical transitions important for the experiments reported in this
paper.  Not to scale.}
\label{fig2}%
\end{figure}

A schematic diagram of the experimental apparatus is shown in Fig. 1, and
the main optical excitation transitions used in the present
experiments are shown in Fig. 2. In Fig. 1, the main focus of the
instrumentation is an optical dipole trap formed in the focal
region of a carbon-dioxide ($CO_{2}$) laser beam. A 40 MHz acousto
optical modulator (AOM) is used to switch and to direct
approximately 50 W of the 100 W $CO_{2}$ laser output to the
trapping region. The trapping beam is focussed to a radial spot
size of $\sim$ 55 $\mu m$, and an associated Rayleigh range of
$z_{R}$ $\sim$ 750 $\mu m$. Because the operating trap wavelength
of 10.6 $\mu m$ is far longer than those of the Rb resonance
transitions, the dipole force is closely proportional to the
static dipole polarizability of atomic Rb; the trap is then
referred to as a quasistatic electric dipole trap (QUEST)
\cite{Grimm}. The trap depth is approximately 650 $\mu K$, and has
associated measured trap angular frequencies $\omega_{z}$ = $2\pi
\cdot 50$ rad/s and $\omega_{r}$ = $2\pi \cdot 1300$ rad/s. As
described in detail elsewhere \cite{parametric1}, the QUEST is
loaded with atoms that have been collected from a surrounding
thermal Rb vapor and cooled in an overlapping magneto optical trap
(MOT). The QUEST atoms are initially loaded from the MOT into the lower energy
F = 1 hyperfine component of the Rb ground level. Following QUEST
loading, the MOT lasers and magnetic fields are shut off, and the
atoms in the dipole trap collisionally thermalize to a temperature
of approximately 65 $\mu$K.  After this natural thermalization
process, approximately 15 $\%$ of the atoms originally in the MOT
have been transferred to the QUEST; this amounts to around 5 million atoms. Measurements of the QUEST
characteristics, after the hold period, by absorption imaging,
parametric resonance, and the measured number of atoms transferred
show a sample with peak density about 5 $\cdot$ 10$^{13}$
$atoms/cm^{3}$ and a temperature of $T_{o}$ = 65 $\mu K$. The transverse Gaussian radius is $9.6 \mu m$, while the longitudinal radius (the Rayleigh length) is $230 \mu m$.  The 1/e lifetime of the confined atoms is greater than 5 s, limited by
background gas collisions.

The primary goal of the experiments is to study light scattering
on the nearly closed $F = 2 \rightarrow F' = 3$ hyperfine
transition associated with the D2 resonance line. However, the
atoms are initially loaded into the QUEST in the F = 1 lower
energy hyperfine component, and must be optically pumped into the
higher energy F = 2 level.  This is accomplished by the MOT
repumping laser beams, tuned resonantly to the $F = 1 \rightarrow
F' = 2 $ transition as indicated in Fig. 2.  As described elsewhere \cite{parametric1} the repumper laser is
an external cavity diode laser, and is locked to an $^{87}$Rb
saturated absorption feature associated with the $F=1\to F'=2$
hyperfine transition. The laser bandwidth is approximately 0.5
MHz. The repumper delivers a beam of maximum intensity $\sim$ 4
$mW/cm^{2}$ and is directed along the same optical paths as the
trapping laser beams. Repumper switching is controlled with an
acousto optical modulator. There are three counter propagating
pairs of such beams directed towards the sample along three
orthogonal directions. The repumper intensity of $\sim$ 4
$mW/cm^{2}$ corresponds to an on-resonance saturation parameter
larger than unity.  With reference to Fig. 1, the light scattered
out of the repumper beams by the atom sample is collected at an
angle of 45 degrees from the horizontal pair of repumper beams and
at 90 degrees from the vertical beam (not shown).
Approximately 1 $\%$ of the resulting scattered light is collected
with a field lens and launched into a multimode fiber which, in
turn, transports the light to an infrared sensitive and
refrigerated photomultiplier tube (PMT) operating in a single
photon counting mode.  The photon counting pulses are amplified
with a fast amplifier and then binned in a multichannel scalar
(MCS) with a time resolution of 5 ns.  Depending on the counting
rate, a data record of several thousand experimental realizations
is necessary to obtain sufficient counting statistics.  Each experimental realization requires several seconds for formation and interrogation of the atomic sample.

\begin{figure}[htpb]
\includegraphics{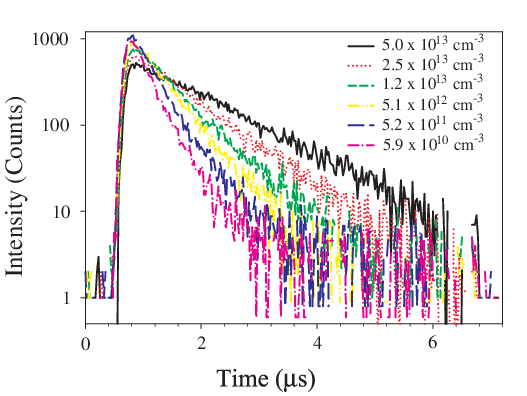}%
\caption{Time-dependence of the light scattered during the F = 1
$\rightarrow$ F' = 2 optical pumping process.  From upper to
lower, the curves correspond to decreasing atomic density; the
values refer to the peak density at the center of the sample. The
correspondence between the peak density, the transverse optical
depth, or the sample dimensions can be found in Table 1.}
\label{fig3}%
\end{figure}

We emphasize that the hyperfine optical pumping
process, by which the $^{87}$Rb atoms in the QUEST are brought
from the F = 1 to the F = 2 ground hyperfine component is an important part of the sample preparation process.
Fluorescence signals taken over a three order of magnitude range
of sample density are shown in Fig. 3.  These densities, as
indicated in the figure legend, represent the peak density in the
central region of the atomic sample. As described previously, the sample density is reduced from its
maximum value by allowing for a selected period of free expansion
of the sample prior to switching on the repumper laser and
collecting the resulting scattered light signals. Further
expansion of the sample during the data acquisition period is
negligible.  Several features of the data are apparent from Fig.
3.  First, there is a small background signal of a few counts
remaining even after the transient signals have decayed away. This
behavior is more apparent in the lower density data, which has an
associated faster decay rate than the higher density data.  This
signal is due to hot atom fluorescence excited by the relatively intense
repumper laser beams. Second, with the background level accounted for, the decay of the signals is well
approximated by a single exponential time decay, with the higher
density data decaying at a much slower rate than the lower density
measurements display. The connection between the various values of
$b_{t}$ and the peak atom density can be found in Table 1.  The time scale associated with the optical
pumping process may be estimated by considering that the repumper
laser beam is intense enough that it saturates the repumper
transition. Then the time scale for the pumping process
transferring an atom from the lower (F=1) to the upper (F = 2)
hyperfine component is approximately $2\tau_o$, where $\tau_o$
$\sim$ 26 ns is the radiative lifetime of the excited state.  On
the average, then the repumper beam penetrates into the sample a
distance of one optical mean free path
$l=\frac{\emph{1}}{\rho\sigma}$ in this time. Here $\rho$ is the
average atom density and $\sigma$ is the optical scattering cross
section \cite{QUESTFormulas} For a sample of size $r_{o}$, an
estimate of the pumping time is then T $\sim$
$2\tau_o\frac{r_{o}}{\emph{l}}$.  This simple formula provides
estimates in fair agreement with the results of Fig. 3.

We point out that in spite of the apparently simple
phenomenology of the optical pumping dynamics, the physical
process is in reality quite complex.  For example, the atom sample
is initially in the lower energy F = 1 hyperfine level, and after
the process is terminated, is nearly completely pumped into the F
= 2 level. At the same time, the atom sample is initially very
optically deep to the repumper laser and optically thin to the F =
2 $\rightarrow$ F$^\prime$ = 2 inelastic Raman decay.  These roles
are reversed but in a spatially inhomogeneous way as the pumping
dynamics take place. Beyond this, multiply scattered light on both
main transitions should participate significantly in the entire
dynamics (this is ignored in the estimate of the pumping time
above).  The entire process is made yet more complex by
the inelastic components in the scattered light, these being
generated by the intense repumper laser.  As these processes are not
the main focus of the present work, we defer theoretical modeling
and more comprehensive experiments to a later study.

To close this section, in the main experimental protocol (see Fig. 1) used to obtain the results reported in the following sections, a probe beam tuned in the
spectral vicinity of the $F = 2 \rightarrow F' = 3$ nearly closed
transition is directed towards the sample, and the resulting
scattered light signals collected by the same optical arrangement
described above. The probe laser is of the same design as the
repumper laser, has a bandwidth $\sim$ 3 MHz, and is switched and
directed by an acousto optical modulator towards the sample.
Because of constraints on the vacuum chamber geometry, the
linearly polarized probe beam is directed (see Fig. 1) at an angle
of approximately 30 degrees away from the fluorescence collection
direction.   The probe beam is also directed downwards at an angle
of 30 degrees. The collection and electronic accumulation of
scattered light signals is the same as with the repumper signals.

Finally, we point out that in some of the
experiments reported here the atomic density was varied over a
wide range.  This was accomplished by allowing for a period of
ballistic expansion of the cloud after the QUEST was turned off.
The atomic sample temperature is known, so this procedure allows
the peak density or the peak optical depth to be determined.   As
the sample is well approximated by a two-axis Gaussian atom
distribution \cite{QUESTFormulas}, the two Gaussian radii and the
peak atom density (or the total number of atoms in the sample),
are sufficient to determine the two peak optical depths
characterizing the sample. We summarize in Table 1 the peak
transverse optical depth $b_{t}$, the peak atom density at the
center of the sample $n_{o}$, the transverse Gaussian radius
$r_{o}$, and the longitudinal Gaussian radius $z_{o}$.  The
optical depth refers here to that of the nearly closed $F = 2
\rightarrow F' = 3$ hyperfine transition, and is obtained by using
a resonance scattering cross section of $1.36 \times 10^{-9}
cm^{2}$.  The optical depth dependence may readily be rescaled to
other $F \rightarrow F'$ transitions \cite{QUESTFormulas}.

\begin{table}
\begin{tabular}{|c|c|c|c|}
  \hline
  % after \\: \hline or \cline{col1-col2} \cline{col3-col4} ...
  Peak $b_{t}$ & $n_{o}$ (atoms/cm$^{3})$ & $r_{o}$ ($\mu m$)& $z_{o}$ ($\mu m)$\\
  \hline
  165 & 5.0 $\times 10^{13}$& 9.8 & 248 \\
  117 & 2.5 $\times 10^{13}$ & 13.8 & 248 \\
  82 & 1.2 $\times 10^{13}$& 19.5 &  248 \\
  53 & 5.1 $\times 10^{12}$& 30.4 & 249 \\
  16 & 5.2 $\times 10^{11}$& 92.3 & 264\\
  5  & 5.9 $\times 10^{10}$& 240 & 345 \\
  \hline

\end{tabular}
\caption{QUEST parameters relating the peak transverse optical
  depth on the $F = 2 \rightarrow F' =3 $ transition
  to the peak sample density and the Gaussian radii of the
  atomic cloud.}
\end{table}

\section{Results: Probe Light Scattering on the $F = 2 \rightarrow F' = 3$
Transition }

\subsection{Density Dependence}
We first turn our attention to the broad main focus of this report, which
is the situation where the hyperfine optical pumping process is
completed prior to initiation of probe light scattering in the
spectral vicinity of the $F = 2 \rightarrow F' = 3$ transition. In
this case, the function of the hyperfine optical pumping is to
transfer the atoms to the F = 2 ground level hyperfine component;
when this process is complete, the peak resonance optical depth on
the probe transition is about 165, as indicated in Table 1.  We
first consider the resonance response as a function of atom
density under weak-field probe conditions of $\sim$ 630 $\mu
W/cm^{2}$. For these measurements, the atom sample is
exposed to a nearly rectangular temporal pulse of 2 $\mu$s
duration; this pulse has a 20 dB rise and fall time of about 100
ns. As shown in Fig. 4(a) and 4(b), the scattered light transients
for all densities consist of a rather rapid increase to an approximate steady
level, followed by a several hundred nanosecond temporal decay
after the probe laser beam is extinguished.

\begin{figure}[htpb]
\includegraphics{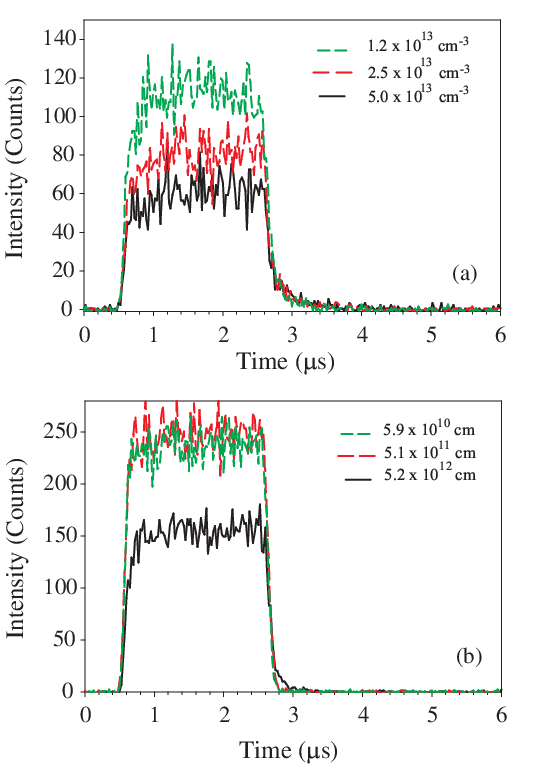}%
\caption{Variations of the $F = 2 \rightarrow F' = 3$ resonant
probe response with density and time.  Note that the maximum
intensity decreases with increasing density, at constant atom
number.  Note also the emergence of a long-time decay after the
probe excitation is shut off.  (a) Dependence for higher
densities. (b) Dependence for lower intensities.}
\label{fig4}%
\end{figure}

We analyze these results by considering the integrated
probe signal (as in Fig. 5) as a function of decreasing atomic
density.  However, instead of the atomic density, we parametrize
the dependence in terms of the peak transverse optical depth
$b_{t}$ through the center of the ellipsoidal atomic sample. The
optical depth is the natural parameter to describe many
characteristics of light scattering and diffusion in dense
scattering media.
Note that there are nominally two optical depths required to
describe our atomic sample \cite{QUESTFormulas}.  These are the
peak transverse optical depth $b_t$, as just mentioned, and the
longitudinal optical depth $b_{l}$, which is typically more than
10 times larger than $b_{t}$. The variation of the integrated
scattered light signals with $b_{t}$ is shown in Fig. 5, where it
is seen that the signals increase as the optical depth decreases.  We point out that the data in Fig. 5 have been previously reported \cite{ScalingJMO} and compared to theoretical results; we have provided additional experimental details here, and include the results for completeness of the present manuscript.
The behavior evident in Fig. 5 can be physically understood from the fact that for
very large $b_{t}$, only the atoms near the sample surface
contribute significantly to the scattering signal. However, as the density, and
correspondingly the optical depth, decreases, more of the atoms in
the sample participate in the scattering, and the resulting signal
increases.  This behavior is a clear indicator that light scattering from these dense and cold atomic samples is a collective process.

The solid curve in Fig. 5 is a theoretical result obtained through approximate scaling laws obtained previously \cite{ScalingJMO}. Those scaling relations were obtained by exploring numerically the dependence of the scattering cross-sections on sample size.  Such scaling rules are useful because the samples explored experimentally contain several orders of magnitude more atoms, making direct numerical simulations impractical. The details of the theoretical approach has been laid out and applied in several previous papers \cite{KupriyanovLightStrong,ScalingJMO,OpticsSpec1,JETP1}, and are briefly summarized in an appendix.

\begin{figure}[htpb]
\includegraphics{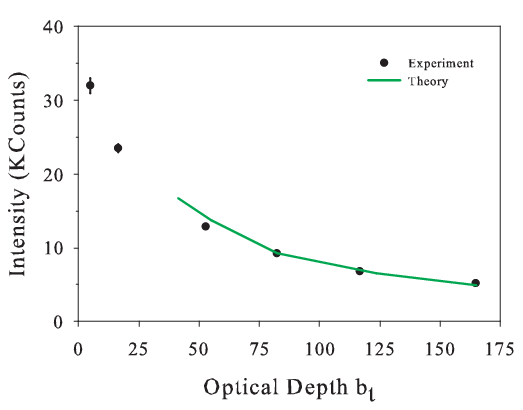}%
\caption{Variations of the integrated $F = 2 \rightarrow F' = 3$
resonant probe response with peak transverse optical depth.  Note
that the maximum intensity decreases with increasing optical
depth, at constant atom number.}
\label{fig5}%
\end{figure}

\subsection{Detuning Dependence}
We now turn to the detuning and time-dependence of the light scattering signals.  In the experiments we have measured the temporal response of the sample to a 2 $\mu s$ probe pulse for detunings $\Delta$ in a $\pm$ 24 MHz range around the atomic $F = 2 \rightarrow F' = 3$ resonance.  The finest temporal resolution is 5 ns, but grouping the data into larger bins of 25 ns, 50 ns or 100 ns width, which we do for the results reported here, improves the signal to noise significantly.  Positive values of $\Delta$ correspond to probe frequencies larger than the isolated atom resonance frequency.

Experimental results are compared to theoretical ones obtained for atomic samples of nearly the same peak density, but fewer atoms.  The theoretical approach used has been described elsewhere \cite{KupriyanovLightStrong,ScalingJMO,OpticsSpec1,JETP1}, and for the convenience of the reader is summarized in the appendix.  To have possibilities to contrast results of the theory with experiments, we choose the density of our motionless four levels atoms in such a way that photons would have the same mean free path as in the $^{87}$Rb samples. Estimating the resonant cross section of the light from a single atom with $J=0\rightarrow J=1$ transition as $3\lambda^2/2\pi$ we obtain that $n_0\simeq 5\cdot\,10^{13}\,cm^{-3}$ in experiment corresponds to $n_0\simeq 0.05\,k^{3}$ in theory. Here $k$ is the wave number of the scattered light.

\subsubsection{Frequency-Dependent Time Response}
Representative measurements of the time response of the light scattered from the sample are given in
Fig. 6, where we see that the response to the nearly rectangular probe pulse consists of a quite rapid increase in
signal to a nearly steady value (within the measurement statistics), followed by a clear decaying signal extending, in the case of excitation around bare atomic resonance, for several $\mu s$ after the probe beam is extinguished.  The time resolution of this data is 25 ns.  The overall features of the results are qualitatively reasonable, for in these high optical depth samples, multiple scattering of light is a strong effect.  Nearer atomic resonance the overall effect of multiple scattering is greater, and is to slow significantly both the build up and the decay of the atomic fluorescence signals.  These data are recorded under conditions of the peak transverse optical depth $b_t$ = 165.  The corresponding theoretical results are shown in Fig. 7, where very good qualitative agreement is seen in comparison with the measurements of Fig. 6.

\begin{figure}[htpb]
\includegraphics{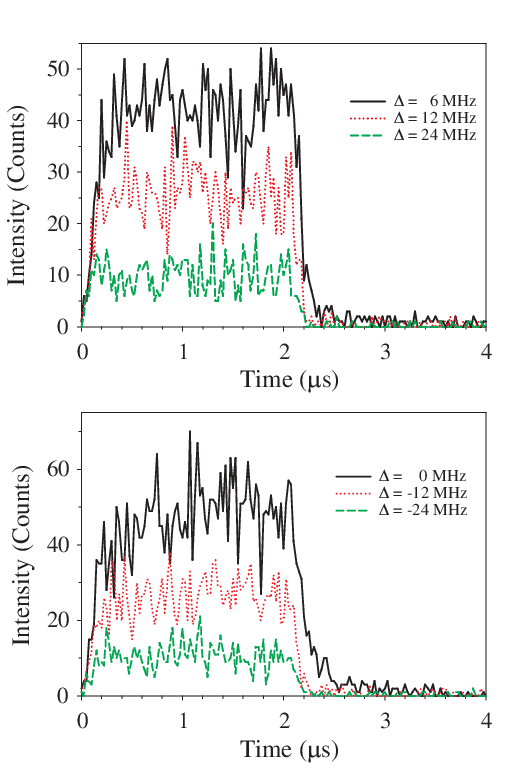}%
\caption{Representative spectral variations of the transient light
scattering response associated with the F = 2 $\rightarrow$ F' = 3
probe transition. Positive (higher frequency) detunings are shown
in (a), while negative (lower frequency) detunings are shown in
(b). These data are recorded under conditions of the peak
transverse optical depth $b_t$ = 165. The bin size on the time axis is 25 ns.}
\label{fig6}%
\end{figure}

\begin{figure}[htpb]
\includegraphics{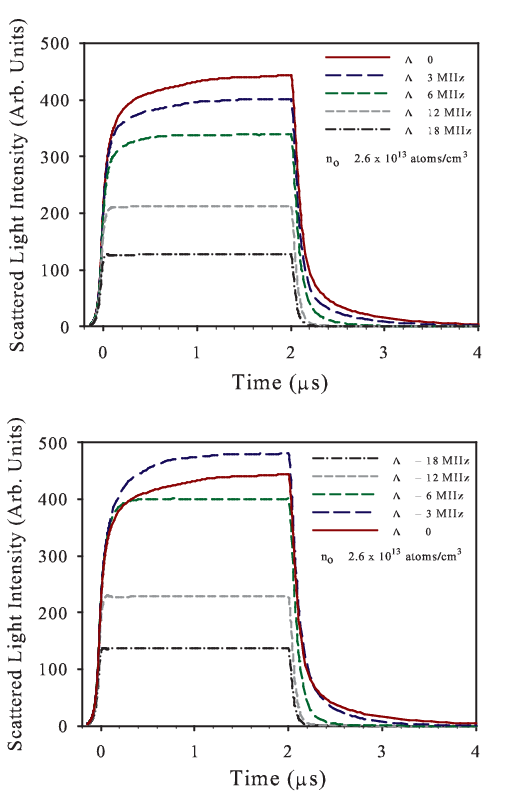}%
\caption{Theoretical spectral variations of the transient light
scattering response associated with the F = 2 $\rightarrow$ F' = 3
probe transition. Positive (higher frequency) detunings are shown
in the upper panel, while negative (lower frequency) detunings are shown in
the lower panel.  The $\Delta = 0$ result is included in each graph for convenience. These results are calculated for
a peak density of $n_o$ = 0.05.}
\label{fig7}%
\end{figure}

\subsubsection{Time-Dependent Spectral Response}
We further analyze the data of Fig. 6 by extracting the response of the system at fixed times but as a function of detuning of the probe frequency from $\Delta$ = 0.  This procedure yields an excitation, or action spectrum, and reveals how strongly probe light of a given detuning yields a response of the atomic sample..  The analysis is largely done with a time resolution of 100 ns in order to obtain improved signal to noise in comparison with smaller bin sizes.  However, for the short period during turn on of the probe pulse, we use a smaller grid of 50 ns, in order to better resolve the relatively rapid changes in the spectrum during this period.  The result of that analysis for the early time part of the turn on spectrum is shown in Fig. 8.  There we can see that the initial width is very large in comparison with the approximate 6 MHz natural width of the bare atom transition.  In the figure, the solid lines through the data points represent fits of a Lorentzian line shape to the measured spectrum.   This choice is not essential, though the correlation of the fit to the data is quite high.   The full width of the spectral response, which is the main extracted quantity, is within a few percent for either Lorentzian or Gaussian fits.  What is clear is that the spectral width of the excitation spectrum is very broad for short times, and narrows quickly.  A similar behavior is seen in the theoretical results presented in Fig. 9, and again, the qualitative agreement is quite good.  We can broadly understand the short-time behavior by realizing that for time scales on the order of the Wigner delay time \cite{Wigner}, the scattered light signal comes from those atoms very near the surface of the atomic cloud.   If we take that thickness to extend until the optical depth b $\sim$ 1, corresponding to a decrease in the probe intensity by $e^{-1}$, then this occurs for a detuning of the probe frequency such that $\Delta_o$ = $\gamma /2$$\sqrt{b_o - 1}$, where $b_o$ is the on resonance transverse optical depth through the center of the cloud.  For our experiment $b_o$ = 165, giving a full width of $\Delta_o$ = $\gamma$$\sqrt{b_o - 1}$ $\sim$ 77 MHz.   This is in very good agreement with the width at very short times (see Fig.8).

\begin{figure}[htpb]
\includegraphics{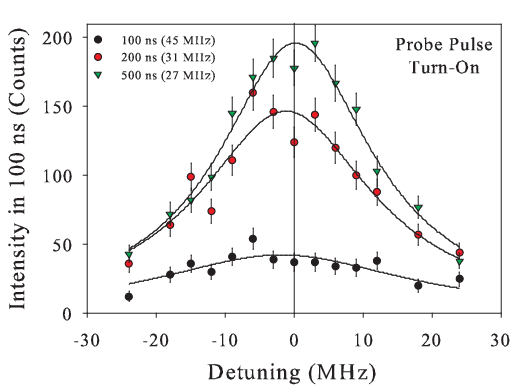}%
\caption{Time dependence of the spectral response of the atomic sample during the turn-on of the weak field probe excitation as a function of probe laser frequency detuning from bare atomic resonance.  These data are recorded under conditions of the peak
transverse optical depth $b_t$ = 165.}
\label{fig8}%
\end{figure}

\begin{figure}[htpb]
\includegraphics{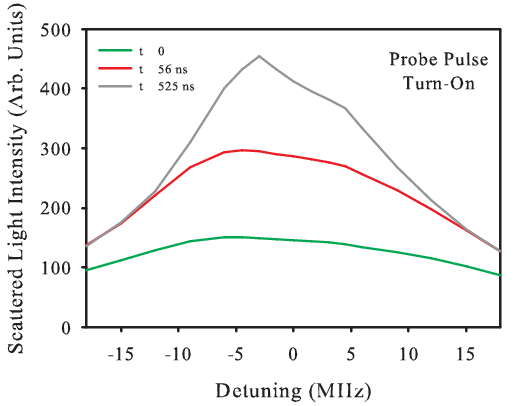}%
\caption{Calculated time dependence of the spectral response of the atomic sample during the turn-on of the weak field probe excitation as a function of probe laser frequency detuning from bare atomic resonance.  These results are calculated for
a peak density of  $n_o$ = 0.05.}
\label{fig9}%
\end{figure}

Upon turn off of the probe pulse the spectral width of the excitation spectrum decreases from a value several times the natural width to approximately 8 MHz.   The experimental results illustrating this are presented in Fig. 10, while corresponding theoretical results are shown in Fig. 11.  As in the spectral response upon turn-on of the prober pulse, the line shape here is also well fit by a Lorentzian form. It is important to note that the spectral width of the probe laser itself is about 3 MHz, and has a measured Gaussian power spectrum.  This means that the widths determined by these measurements are slightly larger than that determined by the physical processes involved.    We also note that there appears to be a small shift of the resonance response to frequencies lower than the single atom resonance. The frequency shift of about -0.4 (4) MHz suggested by the data is of the right magnitude to correspond to the well-known Lorentz-Lorenz (local field) shift at these densities \cite{Maki}. However, further measurements at higher density and with a spectrally narrower probe would be necessary to quantitatively examine the size of the shift.

\begin{figure}[htpb]
\includegraphics{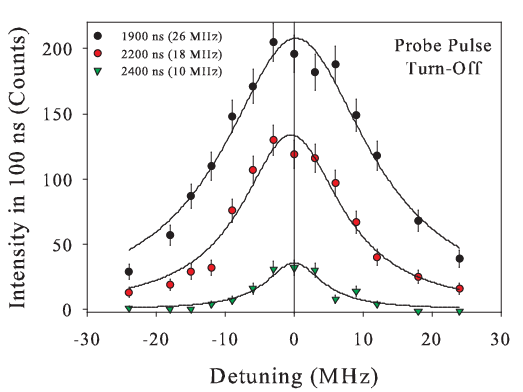}%
\caption{Time dependence of the spectral response of the atomic sample during the turn-off of the weak field probe excitation as a function of probe laser frequency detuning from atomic bare resonance.  These data are recorded under conditions of the peak transverse optical depth $b_t$ = 165.}
\label{fig10}%
\end{figure}

\begin{figure}[htpb]
\includegraphics{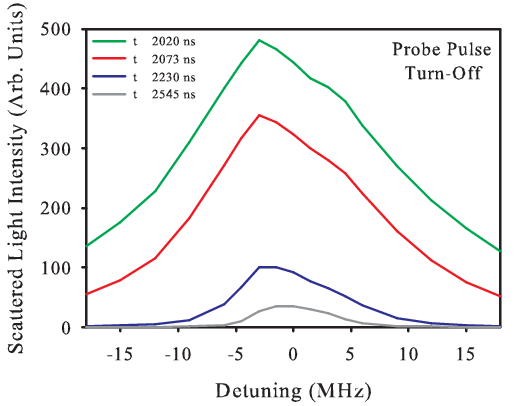}%
\caption{Calculated time dependence of the spectral response of the atomic sample during the turn-off of the weak field probe excitation as a function of probe laser frequency detuning from bare atomic resonance. These results are calculated for
a peak density of  $n_o$ = 0.05.}
\label{fig11}%
\end{figure}

We now turn to the excitation spectra for the full temporal range of the data; the experimental results are presented in Fig. 12.   There we see the prompt narrowing, as mentioned before, of the spectral width of the excitation spectrum to a nearly steady level of around 20 MHz in about 500 ns.   This approximately steady state level persists until the probe is shut off, when there occurs a second fairly sharp decrease to a level just a bit larger than the 6 MHz natural line width of the resonance transition.  Similar results are seen in the calculation of the full spectrum, as shown in Fig. 13.  We note that the spectral width of the action spectra decreases slightly through most of the time period during which the probe pulse is on.  We attribute this effect to the fact that for off resonance excitation the effective optical depth is quite a bit less than for near resonance excitation.   This means that it takes longer for steady state in the scattered light intensity to be reached for near resonance excitation. This behavior is also evident in Fig. 6, where the on-resonance intensity is increasing for the first $\mu$s or so of the signal.   As the approach to steady state leads to an increased intensity near resonance, the effective full-width at half maximum would be expected to decrease slightly, as it does.

\begin{figure}[htpb]
\includegraphics{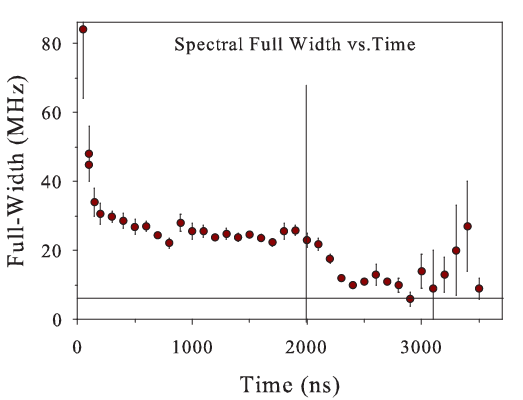}%
\caption{Time dependence of the spectral response of the atomic sample to weak field probe excitation at different detunings from atomic bare resonance.  These data are recorded under conditions of the peak
transverse optical depth $b_t$ = 165.}
\label{fig12}%
\end{figure}

\begin{figure}[htpb]
\includegraphics{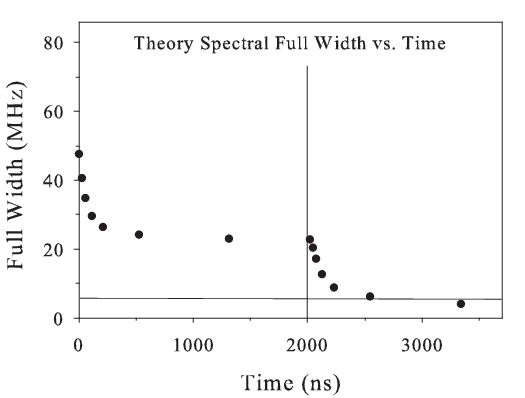}%
\caption{Theoretical calculations of the time dependence of the spectral response of the atomic sample to weak field probe excitation at different detunings from atomic bare resonance.  These results are calculated for
a peak density of  $n_o$ = 0.05.}
\label{fig13}%
\end{figure}

We may gain some understanding of the experimental and theoretical results by recalling that the light scattering amplitudes may be determined by the poles of the projected resolvent matrix, as defined in the Appendix.   In each realization of the atomic sample, for which the locations of the atoms are generally different, there is a set of poles, each consisting of a real and imaginary part; these determine, respectively, the spectral location and the width of a resonance associated with a collective mode of excitation of the sample.   For many realizations of the sample, a distribution of such collective states are found. Distributions for the real and imaginary parts of the poles are shown in Figs. 14 - 15.  Plotted on the ordinate in each case is the number of states obtained normalized to the number of atoms.  We note that dividing this quantity by the energy bin size in each case would result in a density of states.  For the real parts of the poles (Fig. 14) this quantity is 0.1 $\gamma$ = 600 KHz.   For the imaginary parts of the poles (Fig. 15), the scale is 0.0025 $\gamma$ = 15 KHz.  Note also that the number of states for the real parts of the poles are grouped according a range of decay rates, as indicated in the legend of Fig. 14.   Similarly, in the legend of Fig. 15 the grouping of the number of states for the imaginary parts of the poles is given in terms of different detuning ranges.  This grouping allows us to make qualitative insights into the experimental behavior seen in Figs. 6, 8, 10, and 12, and the theoretical results of Figs. 7, 9, 11, and 13.

In considering the results of Fig. 12, we first reiterate that the initial response of the system consists of collective scattering from the atomic sample as a whole.   This signal constitutes a very small part of the total scattered intensity at any time, and is observable at only during the probe pulse turn-on period, a time interval on the order of the 50 ns scattering time \cite{Wigner}, see Fig. 6.  It is spectrally broad and weak because the signals arise from a surface layer of about one absorption length in thickness and thus only a relatively small number of the atoms in the sample contribute.  Multiple scattering becomes increasingly important at longer times, and steady state closely reached after a few hundred nanoseconds.  Here the probe laser excites a distribution of quasimodes \cite{Cooperative} having a characteristic resonance frequency and decay rate which correspond to the real and imaginary parts of the resolvent poles.   In the steady state mode excitation is balanced by mode decay and the nearly constant spectral width of the excitation spectrum reflects this.  Once the probe laser is extinguished, the width of the excitation spectrum rapidly decreases.   This is a reflection of the mode distribution having different lifetimes, as seen in Fig. 14.   There the distribution of states is binned according to different decay rate bands, as defined in the legend.   We see that the distribution of pole real parts indeed narrows as the modes of larger $\Gamma$ die away.

Last, we turn to the longest time decay evident in Fig. 12 as the longer tails appearing after the probe pulse has been turned off and the shorter lived transients have died away.  These transients have an associated lifetime of about 650 (50) ns, and are independent of detuning over the range $\pm$ 6 MHz (reliable results could not be obtained for larger detunings).  This is expected, as the longest lived mode in the sample, the so-called Holstein mode, has a characteristic lifetime, but it is excited with lower efficiency at larger detunings.   Inspection of the calculations in Fig. 15 shows the distribution of pole imaginary parts for different detuning bands.  The graphs show that for detunings $\Delta$ larger than $\pm$ 6 MHz ($\pm$ $\gamma$), the longest lived modes (smallest $\Gamma$) are indeed suppressed. The longest lived mode here implies a statistical value, because each realization of the atomic sample has a different spatial distribution of atom.  In Fig. 15, this is the peak of the 'all states' distribution, and shows that for this particular average atom distribution the peak corresponds to a decay rate of about 0.025$\gamma$, and an associated lifetime of about 1 $\mu$s.  Considering the physical differences between the model sample and the experimental one, as discussed earlier, this is in qualitative agreement with the experimental value of 650 (50) ns.

\begin{figure}[htpb]
\includegraphics{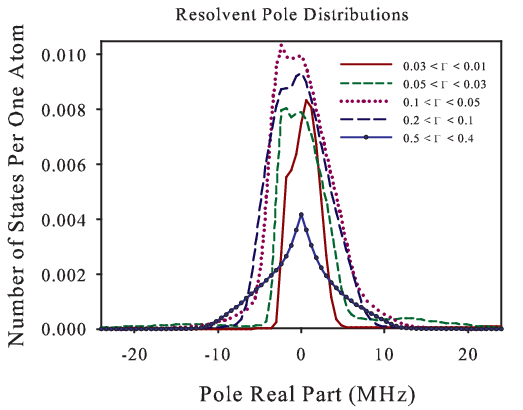}%
\caption{Distribution of the number of states per atom as a function of the real part of the pole of the resolvent. Results are shown accumulated within the indicated range of imaginary parts of the poles of the resolvent.  The quantity $\Gamma$ scales the relaxation rate associate with each pole according to 1/$\Gamma$ = 26.24 ns. These results are calculated for a peak density of $n_o$ = 0.05.}
\label{fig14}%
\end{figure}

\begin{figure}[htpb]
\includegraphics{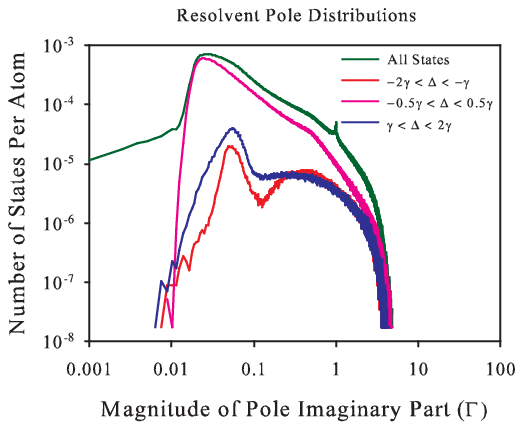}%
\caption{Distribution of the number of states per atom as a function of the imaginary part of the pole of the resolvent. Results are shown accumulated within the indicated range of imaginary parts of the poles of the resolvent.  Note that all pole imaginary parts are negative; the magnitudes are plotted here.  The quantity $\gamma$ scales the detuning range associated with each pole according to $\gamma$ = 6 MHz. These results are calculated for  a peak density of $n_o$ = 0.05.}
\label{fig15}%
\end{figure}

\section{Conclusions}
In this paper we have reported studies of near resonance light scattering from dense and ultracold atomic rubidium. The density is sufficiently large that the experimental conditions are close to satisfying the Ioffe-Regal criterion for light localization on the optically closed $F = 2 \rightarrow F' = 3$ transition.  We have observed collective effects associated with the density and detuning dependence, and suggestion of the influence of the dipole dipole interaction between the Rb atoms.  We have also observed evolution of the excitation spectral distributions towards spectrally narrow features indicative of long lived modes within the sample.  Results are also presented of theoretical treatments of the quantities determined in the experiments; very good correspondence between the experimental and theoretical results is obtained.  Although the experimental results show collective effects, there was no evidence of cooperative effects, or clear evidence of very long lived states which might be associated with Anderson localization of light in three dimensions.   The experimental results are then understood, by comparison with theoretical calculations, in terms of multiple scattering at low temperatures where the effect of atomic velocity is minimal, and the time scales for effects associated with atomic recoil have not been attained.  We conclude that, for the $F = 2 \rightarrow F' = 3$ transition, the large optical depth of the sample does not permit efficient injection of optical excitation to the highest density central portion of the sample.  Observation of localization effects then seems difficult using this scheme.   One alternative is to use light shifts to manipulate the optical depth to achieve optical control of light injection into the samples; such exploration is now underway.

\section{Acknowledgments}
We appreciate financial support of the National Science Foundation (Grant Nos. NSF-PHY-0654226 and NSF-PHY-1068159), the Russian Foundation for Basic Research (Grant No. RFBR-CNRS 12-02-91056). D.V.K. would like to acknowledge support from the External Fellowship Program of the Russian Quantum Center (Ref. Number 86). We also acknowledge the generous support of the Federal Program for Scientific and Scientific-Pedagogical Personnel of Innovative Russia for 2009-2013 (Contract No. 14.B37.21.1938).

\section{Appendix. Theoretical Approach}
In this section we provide a sketch of the approach we have used for the theoretical results presented in this paper.  In this sketch, we follow the descriptions in our earlier papers \cite{JETP1,KupriyanovLightStrong}  We consider the temporal dynamics of a system consisting of $N+1$ motionless atoms. $N$ atoms form the cloud. These atoms are identical and have a ground state of angular momentum $J=0$ separated by the energy $\hbar \omega_a$ from an excited $J=1$ state. The decay constant of this state is taken as $\gamma$. One atom plays the role of the light source. It is located far from the cloud and has the same $J=0 \leftrightarrow J=1$  structure of levels but a different transition frequency $\omega_s$ and a different decay constant $\gamma_s$. We  assume that initially all atoms of the cloud are in the ground state and the spatially separated source atom is in a coherent state which is a superposition of the ground and a small admixture of the excited state.

The system dynamics can be described by a non stationary Schrodinger equation for the wave function $\psi $ of the joint system consisting of atoms and the field generated in the process of the evolution
\begin{equation} i\hbar \frac{\partial \psi
}{\partial t}=H\psi .  \label{1}
\end{equation}
The atom-field interaction is described in the dipole approximation
\begin{equation}
V=-\sum_{a}\mathbf{d}^{(a)}\mathbf{E}(\mathbf{r}_{a}),  \label{3}
\end{equation}%
\begin{eqnarray}
\mathbf{E}(\mathbf{r})&=&\mathbf{E}^{(+)}(\mathbf{r})+\mathbf{E}^{(-)}(\mathbf{%
r})=\label{4}\\
&=&i\sum\limits_{\mathbf{k},\alpha }\sqrt{\frac{2\pi \hbar \omega _{k}}{\cal V}}%
\mathbf{e}_{\mathbf{k}\alpha }a_{\mathbf{k}\alpha }\exp (i\mathbf{kr})+h.c.
\nonumber
\end{eqnarray}%
where $\mathbf{E}^{(\pm )}$ are the operators of the positive and negative frequency components of the field;
$a_{\mathbf{k}\alpha }$ is the photon annihilation operator in a mode with wave vector $\mathbf{k}$ and polarization $\alpha ;$ $\cal V$ is the quantization volume; $\mathbf{d}^{(a)}$  is the dipole moment operator of the atom $a$, $\mathbf{e}_{\mathbf{k}\alpha }$ are polarization unit vectors.

We look for a solution of the Schrodinger equation as a superposition of eigenstates  of the
operator $H_{0}$
\begin{eqnarray}
\psi & =&\sum_{\mathbf{k}\alpha}b_{\mathbf{k}\alpha }(t)|g,g,...g\rangle \otimes |\mathbf{k}\alpha \rangle +
\notag  \\&& \sum_{j}b_{j}(t)|g,g,...g,e_j,g,...g\rangle \otimes |vac\rangle+
\notag  \\&& b'_0(t)|g,g,...g\rangle \otimes |vac \rangle+ \label{6}
\\&& \sum_{\mathbf{k}\alpha,j,l}b_{\mathbf{k}\alpha,j,l }(t)|g,...g,e_j,g...g,e_l,g,...g\rangle \otimes |\mathbf{k}\alpha \rangle \notag
\end{eqnarray}
In the rotating wave approximation it is enough to take into account only the two first items of this expression. States without excitation both in the atomic and field subsystem allow us to describe coherent states of the source atom. Non resonant states with two excited atoms and one photon  are necessary for a correct description of the dipole-dipole interaction at short interatomic distances. Note that, in the considered case, there are three excited states for each atom  $e=|J,m\rangle$, which differ by the value of angular momentum projection $m=-1,0,1$.

As an initial condition for Eq. (\ref{1}) we will consider the case when at $t=0$ the field is in a vacuum state, all atoms of the cloud are in the ground state and the source atom, which we denote by index $s$,  is in superposition of ground and one of the excited states $|J,m\rangle$. Designating  the corresponding amplitudes as  $b'_0$ and $b_0$, we can write

\begin{equation}
\psi(0) =(b'_0| g,g,...g \rangle +b_0|e_{s0}, g,g,...g \rangle ) \otimes |vac\rangle,
\label{9}
\end{equation}

\noindent where the index $e_{s0}$ corresponds to the one of the three possible states of atom $s$ which is populated in the initial moment of time.

In the framework of the assumptions made, the amplitude $b'_{0}(t)$  does not change during the evolution of the system $b'_{0}(t)=b'_{0}$, because transitions between this and the other states taken into account are impossible.

To determine all other amplitudes we solve the set of equations which follows from  (\ref{1}). We exclude amplitudes of states with one photon and obtain a finite closed system of equations for $b_{e}(t)\equiv b_{e_{j}}(t);\, j \neq s $. For Fourier components $b_{e}(\omega)$ we have (at greater length see \cite{JETP1})
\begin{equation}
\sum\limits_{e^{\prime }\neq s}\left[ (\omega -\omega _{a})\delta
_{ee^{\prime }}-\Sigma _{ee^{\prime }}(\omega )\right] b_{e^{\prime
}}(\omega ) =\Lambda_{es}(\omega ).
\label{19}
\end{equation}

Matrix elements $\Sigma _{ee^{\prime }}(\omega )$ for $e$ and $e^\prime$ corresponding to different atoms describe excitation exchange between these atoms. Assuming that in state $\psi _{e^{\prime }}$  and $\psi _e$ atoms $b$  and $a$ are excited correspondingly, in the framework of the pole approximation, we have
\begin{eqnarray}
&&\Sigma _{ee^{\prime }}(\omega )=\sum\limits_{\mu ,\nu}
\frac{\mathbf{d}_{e_{a};g_{a}}^{\mu }\mathbf{d}_{g_{b};e_{b}}^{\nu }}{\hbar r^{3}}\times \\&& \left[ \delta _{\mu \nu }\left(
1-i\frac{\omega _{a}r}{c}-\left( \frac{\omega _{a}r}{c}\right) ^{2}\right)
\exp \left( i\frac{\omega _{a}r}{c}\right) \right. -
\notag  \\&&
\left. -\dfrac{\mathbf{r}_{\mu }\mathbf{r}_{\nu }}{r^{2}}\left( 3-3i\frac{\omega _{a}r}{c}-\left( \frac{\omega _{a}r}{c}\right) ^{2}\right) \exp
\left( i\frac{\omega _{a}r}{c}\right) \right] .
\notag
\end{eqnarray}
Here $\mathbf{r}_\mu$ is the projection of the vector $\mathbf{r}=\mathbf{r}_{a}-\mathbf{r}_{b}$ on the axis of the chosen coordinate system and  $r=|\mathbf{r}|$ is the separation between atoms $a$ and $b$.

If $e$ and $e^\prime$ correspond to excited states of one atom then $\Sigma _{ee^{\prime }}(\omega )$ differs from zero only for $e=e^\prime$ ($m=m^\prime$). In this case $\Sigma _{ee}(\omega )$ determines the Lamb shift and the decay constant of corresponding excited state. Including Lamb shifts in the transition frequency $\omega _{a}$ we get
\begin{equation} \Sigma _{ee}(\omega )=-i\gamma _{a}/2.  \label{27}
\end{equation}

The term $\Lambda_{es}(\omega )$ in the right-hand side of Eq. (\ref{19}) describes excitation of the cloud atoms by the radiation of the source.
For modeling the time evolution of cloud excitation in a real experiment (see part \dots above) we assume that decay constant of the source atom is very small $\gamma_s \rightarrow 0$ and at the same time we employ the presence of special shutter between the light source and the cloud. This allows us to simulate the effect of the shape of the experimental probe pulse, which is normally well characterized in experiments.  The time profile of the shutter $F(t)$ is taken as trapezoid. Neglecting the secondary excitation of the source atom $s$ by reradiation from the cloud and assuming that the size of the atomic ensemble is negligible compare with the distance from it to the source, we have
\begin{eqnarray}
\Lambda_{es}(\omega )&=&b_{0}F(\omega)\widetilde{\Sigma}_{es}(\omega ),  \label{27.1}\\ \\
\widetilde{\Sigma} _{es }(\omega )&=&-\sum\limits_{\mu ,\nu}
\frac{k^{2}\mathbf{d}_{e;g}^{\mu }\mathbf{d}_{g_{s};e_{s}}^{\nu
}}{\hbar r_{s}}\left[ \delta _{\mu \nu }-\frac{\mathbf{k}_{\mu
}\mathbf{k}_{\nu }}{k^{2}}\right]\times\nonumber\\ &&\times \exp \left( ikr_{s}+i\mathbf{kr}_{e}\right).
\label{27.2}
\end{eqnarray}
Here $\mathbf{k}=\omega \mathbf{n/}c$; $\mathbf{r}_{e}$ are radii locating the atoms; $\mathbf{n}$ is a unit vector oriented from the source to the cloud and  $F(\omega)$ is a Fourier transform of the excitation time profile $F(t)$.

Knowledge of explicit expressions for $\Lambda_{es}(\omega )$ and $\Sigma _{ee^{\prime }}(\omega )$ allows us to determine the amplitudes of all one-fold excited states.

Introducing the inverse matrix which, as shown in \cite{KupriyanovLightStrong}, is a resolvent operator of the considered system projected on the states consisting of the single atom excitation, distributed over the ensemble, and the vacuum state for all the field modes,

\begin{eqnarray}
R_{ee^{\prime }}(\omega )=\left[ (\omega -\omega _{a})\delta _{ee^{\prime }}-\Sigma
_{ee^{\prime }}(\omega )\right] ^{-1},  \label{21.1}
\end{eqnarray}
we can write the solution of the system (\ref{19}) as follows
\begin{eqnarray}
b_{e}(\omega ) =\sum\limits_{e^{\prime }\neq
s}R_{ee^{\prime }}(\omega )\Lambda _{e^{\prime }s}(\omega ).  \label{21}
\end{eqnarray}
For amplitude $b_{e}(t)$ we get
\begin{eqnarray}
b_{e}(t) &=&\int\limits_{-\infty }^{\infty }\dfrac{d\omega }{2\pi }%
b_{0}\exp (-i\omega t)F(\omega)\sum\limits_{e^{\prime }\neq s}R_{ee^{\prime }}(\omega
)\tilde{\Sigma} _{e^{\prime }s}(\omega ).\nonumber \\ \label{22}
\end{eqnarray}%

This relation gives the distribution of excited states at any instant of time.  Knowing the amplitudes $b_{e}(t)$ allows us to determine all other amplitudes $b(t)$ incoming in Eq.(\ref{6}) and consequently to calculate the properties of scattered light (for details see \cite{JETP1}), particularly the angular distribution, polarization and time evolution of its intensity.

In the present work to determine the time dynamics of the system we calculate the integral like (\ref{22}) by means of the residue theory. The poles of the matrix of the projected resolvent $R_{ee^{\prime }}$ play a key role in the calculation.  These poles are determined in turn by eigenstates of the matrix $\Sigma _{ee^{\prime }}$. Decomposing the vector $\tilde{\Sigma} _{e^{\prime }s}$ over the eigenvector of the matrix $\Sigma _{ee^{\prime }}$ we present the integral (\ref{22}) as the sum of separate pole contributions. The energy and decay constant of each pole depend on both the real and imaginary parts of the eigenvalues of $\Sigma _{ee^{\prime }}$.


\begin{thebibliography}{99}
\bibitem{QUESTFormulas}For an ellipsoidal Gaussian atom
distribution of sizes $r_{o}$ and $z_{o}$ and peak density
$n_{o}$, $n(r) = n_{o}e^{-r^{2}/2r_{o}^{2}-z^{2}/2z_{o}^{2}}$. The
total number of atoms is $N = (2\pi )^{3/2}n_{o}r_{o}^{2}z_{o}$
and the peak transverse (longitudinal) optical depth is
$b_t=\sqrt{2\pi}n_o\sigma_or_o$ ($b_l=\sqrt{2\pi}n_o\sigma_oz_o$).
$\sigma_o$ is the weak field resonance light scattering cross
section.  The peak total cross-section is given by
$\frac{2F^{\prime} + 1}{2F+1}\frac{\lambda^{2}}{2\pi}$.
\end{thebibliography}
\end{document}